\documentclass[twocolumn,twoside,slac_two]{revtex4}
\usepackage{graphicx}
\usepackage{fancyhdr}
\usepackage{hyperref}
\hypersetup{
    colorlinks=true,
    urlcolor=magenta,
}
\begin{document}

\title{The Solar Orbiter Mission: an Energetic Particle Perspective}

\author{R. G\'omez-Herrero}
\author{J. Rodr\'iguez-Pacheco}
\affiliation{Universidad de Alcal\'a, E-28871, Alcal\'a de Henares, Spain}
\author{R. F. Wimmer-Schweingruber}
\affiliation{Christian-Albrechts-Universit\"at zu Kiel, D-24118, Kiel, Germany}
\author{G.M. Mason}
\affiliation{Applied Physics Laboratory, Johns Hopkins University, Laurel, MD 20723, USA}
\author{S. S\'anchez-Prieto}
\affiliation{Universidad de Alcal\'a, E-28871, Alcal\'a de Henares, Spain}
\author{C. Mart\'in}
\affiliation{Christian-Albrechts-Universit\"at zu Kiel, D-24118, Kiel, Germany}
\author{M. Prieto}
\affiliation{Universidad de Alcal\'a, E-28871, Alcal\'a de Henares, Spain}
\author{G.C. Ho}
\affiliation{Applied Physics Laboratory, Johns Hopkins University, Laurel, MD 20723, USA}
\author{F. Espinosa Lara}
\author{I. Cernuda}
\author{J.J. Blanco}
\author{A. Russu}
\author{O. Rodr\'iguez Polo}
\affiliation{Universidad de Alcal\'a, E-28871, Alcal\'a de Henares, Spain}
\author{S.R. Kulkarni}
\author{C. Terasa}
\author{L. Panitzsch}
\author{S.I. B\"ottcher}
\author{S. Boden}
\author{B. Heber}
\author{J. Steinhagen}
\author{J. Tammen}
\author{J. K\"ohler}
\author{C. Drews}
\author{R. Elftmann}
\author{A. Ravanbakhsh}
\author{L. Seimetz}
\author{B. Schuster}
\author{M. Yedla} 
\affiliation{Christian-Albrechts-Universit\"at zu Kiel, D-24118, Kiel, Germany}
\author{E. Valtonen}
\author{R. Vainio}
\affiliation{Department of Physics and Astronomy, University of Turku, 20014, Turku, Finland}

\begin{abstract}
Solar Orbiter is a joint ESA-NASA mission planed for launch in October 2018. The science payload includes remote-sensing and in-situ instrumentation designed with the primary goal of understanding how the Sun creates and controls the heliosphere. The spacecraft will follow an elliptical orbit around the Sun, with perihelion as close as 0.28 AU. During the late orbit phase the orbital plane will reach inclinations above 30 degrees, allowing direct observations of the solar polar regions. The Energetic Particle Detector (EPD) is an instrument suite consisting of several sensors measuring electrons, protons and ions over a broad energy interval (2 keV to 15 MeV for electrons, 3 keV to 100 MeV for protons and few tens of keV/nuc to 450 MeV/nuc for ions), providing composition, spectra, timing and anisotropy information. We present an overview of Solar Orbiter from the energetic particle perspective, summarizing the capabilities of EPD and the opportunities that these new observations will provide for understanding how energetic particles are accelerated during solar eruptions and how they propagate through the Heliosphere.
\end{abstract}

\maketitle

\thispagestyle{fancy}

\section{THE SOLAR ORBITER MISSION}
During the last decades, several space-based observatories have decisively contributed to improve our knowledge of different aspects of the Physics of the Sun and the heliosphere. However, there have been no missions before Solar Orbiter \citep{Muller_2013} and Solar Probe Plus (SPP, \cite{Fox_2016}) specifically conceived to explore the link between the Sun and the Solar wind. Solar Orbiter is a joint science mission between the European Space Agency (ESA) and the National Aeronautics and Space Administration (NASA) designed with the primary objective of understanding how the Sun creates and controls the heliosphere. Solar Orbiter is the first medium-class mission of ESA's Cosmic Vision program. It is currently planned for launch in October 2018, carrying onboard a scientific payload consisting of a comprehensive set of remote-sensing (RS) and in-situ (IS) instruments designed to measure from the photosphere into the solar wind, working together to answer four interdependent top-level science questions:

\begin{itemize}
	\item What drives the solar wind and where does the coronal magnetic field originate from?
	\item How do solar transients drive heliospheric variability?
	\item How do solar eruptions produce energetic particle radiation that fills the heliosphere?
	\item How does the solar dynamo work and drive connections between the Sun and the heliosphere?
\end{itemize}

The third goal above is directly focused on the origin of Solar Energetic Particles (SEP). SEP events are of great interest not only from the Space Weather point of view but also because Solar Orbiter with its comprehensive set of instruments will provide a unique opportunity to understand the physics of the injection, acceleration and escape processes of energetic particles, which is an important and ubiquitous astrophysical process which can only be studied in situ in the heliosphere.

The three-axis stabilized Sun-pointing spacecraft (s/c) will follow an elliptical orbit around the Sun, with perihelion as close as 0.28 AU, allowing IS measurements of the plasma, fields, waves and energetic particles in a region where much of the crucial physics related with solar activity takes place and is relatively undisturbed by interplanetary propagation processes \citep{Muller_2013}. At the same time, high-resolution imaging instruments will permit a direct link between the IS observations and the corresponding solar sources. While former pioneering missions such as Helios \citep{Porsche_1977}, already explored the innermost region of the heliosphere and highlighted the importance of IS measurements close to the Sun, the lack of RS instrumentation onboard did not allow a detailed exploration of the connections between solar structures and IS observations. 

During the late orbit phase the s/c orbital plane will reach inclinations $>$30 degrees above the heliographic equator, allowing direct observations of the solar polar regions. During limited time intervals the s/c orbital period will be close to the solar rotation period (quasi co-rotation), being able to monitor the evolution of the same solar region during extended periods. 

\begin{figure}
\includegraphics[width=80mm]{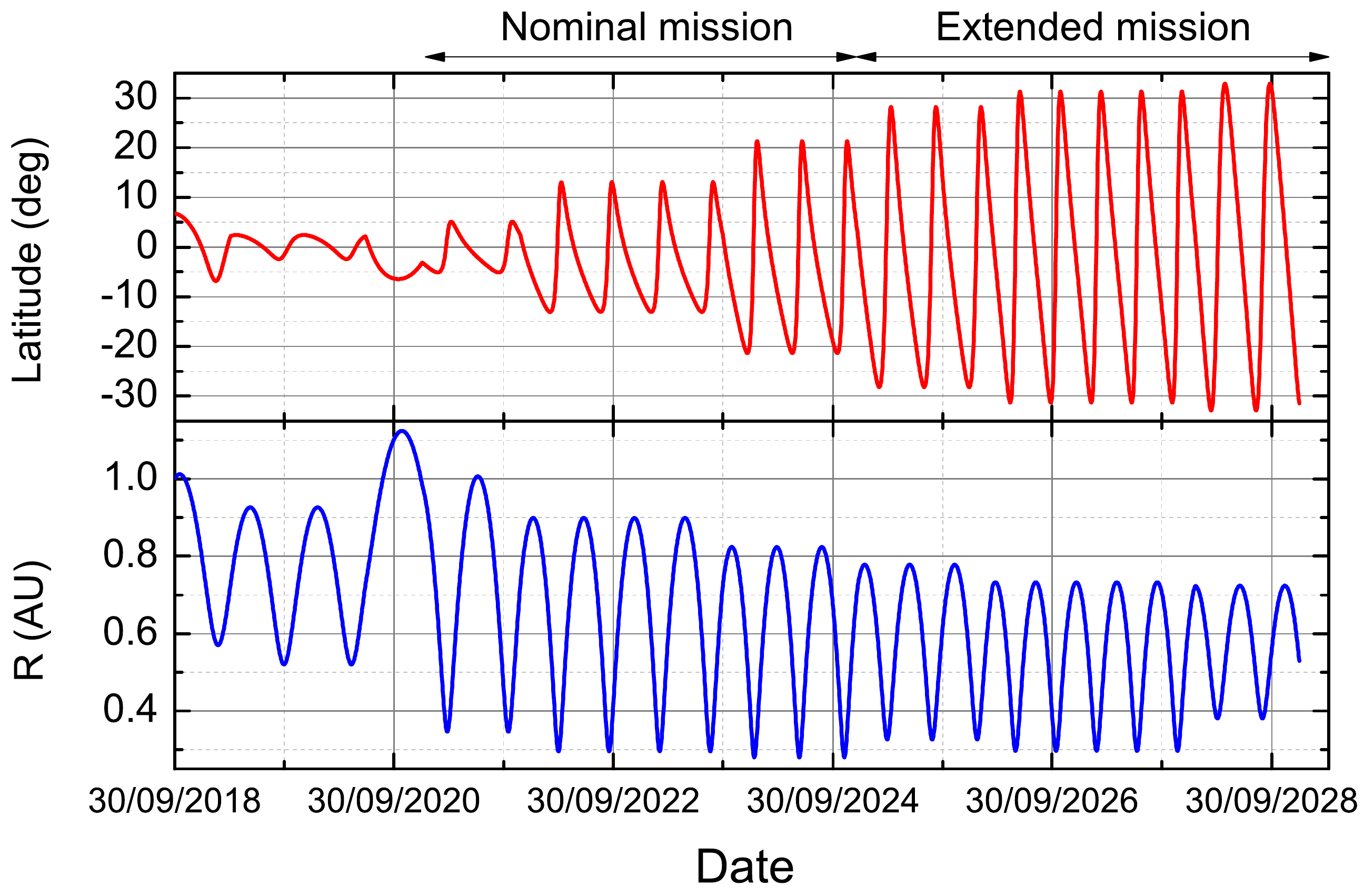}
\caption{Solar Orbiter heliographic latitude (top) and heliocentric distance (bottom) versus time.}
\label{figorbit}
\end{figure}

Solar Orbiter has been optimized in order to reuse designs and technology from the BepiColombo mission \citep{Benkhoff_2010} to Mercury. A sophisticated heat shield will protect the conventional s/c and the payload from the intense direct solar flux when approaching perihelion. The two solar arrays can be tilted in order to control overheating close to the Sun. Multiple planetary gravity assist maneuvers (GAM) at Earth and Venus will be used in order to reach the final elliptical orbit and to gradually increase the orbit inclination. After a Near Earth Commissioning Phase (NECP) and a cruise phase lasting more than 2 years, the s/c will start its 4-year long nominal mission phase (NMP). The first perihelion $<$0.3 AU will be reached 3.5 years after launch (see Figure \ref{figorbit}). During the NMP, the orbit inclination relative to the solar equator will remain below 25 degrees, which will increase up to a maximum of $\sim$33 degrees during the extended mission phase (EMP). With the current mission schedule, the maximum of solar cycle 25 will be covered by the NMP and EMP. Some key-facts of the s/c are summarized in Table \ref{tabscfacts}. 

\begin{table}[t]
\caption{Solar Orbiter s/c key facts.\label{tabscfacts}}
\begin{tabular}{|l|l|}
\hline 
Dimensions &2.5$\times$3.0$\times$2.5 m$^3$\\
&(launch configuration)\\
\hline 
Overall mass &  1750 kg\\
\hline 
Maximum power demand &1100 W\\
\hline 
Launcher &Atlas V 411 (NASA)\\
\hline 
Mission operations center &ESOC, Darmstadt, Germany\\
\hline 
Science operations center &ESAC, Villafranca, Spain\\
\hline 
Nominal ground station &Malarg\"ue, Argentina\\
\hline
\end{tabular}
\end{table}

The 10 instruments that comprise the scientific payload are summarized in Table \ref{tabinstruments}. RS instruments will provide data only during certain time intervals of particular scientific relevance (RS windows), around perihelia and high latitude periods. IS instruments will operate continuously, starting at cruise phase. Instruments can trigger themselves autonomously to higher cadence modes (burst mode) during limited periods of special scientific interest.

\begin{center}
\begin{table*}
\caption{Solar Orbiter scientific payload.\label{tabinstruments}}
\begin{tabular}{|l|l|l|l|}
\hline 
\textbf{Instrument} & \textbf{Acronym} & \textbf{Type} & \textbf{Principal Investigator}\\
\hline 
Energetic Particle Detector suite & EPD & IS &J. Rodr\'iguez-Pacheco (Spain)\\
\hline 
Magnetometer experiment & MAG & IS & T.S. Horbury (UK)\\
\hline 
Radio and Plasma Waves experiment & RPW & IS & M. Maksimovic (France)\\
\hline 
Solar Wind Analyser instrument suite & SWA & IS & C.J. Owen (UK) \\
\hline 
Extreme Ultraviolet Imager & EUI & RS & P. Rochus (Belgium) \\
\hline 
Multi Element Telescope for Imaging & METIS & RS & E. Antonucci (Italy) \\
and Spectroscopy (Coronagraph) & & &\\
\hline 
Polarimetric and Helioseismic Imager (Magnetograph) & PHI & RS & S.K. Solanki (Germany) \\
\hline 
Solar Orbiter Heliospheric Imager & SoloHI & RS & R.A. Howard (USA) \\
\hline 
Spectral Imaging of the Coronal Environment& SPICE & RS & European-led\\
(Extreme ultraviolet imaging spectrograph)& & & facility instrument\\
\hline 
Spectrometer/Telescope for Imaging X-rays & STIX & RS & S. Krucker (Switzerland)\\
\hline
\end{tabular}
\end{table*}
\end{center}

\section{THE ENERGETIC PARTICLE DETECTOR (EPD) SUITE}
\subsection{Key science questions}
The inner heliosphere is filled by various energetic particle populations of diverse origin. These populations include contributions from impulsive and gradual Solar Energetic Particle (SEP) events accelerated during solar eruptive phenomena such as flares and coronal mass ejections (CMEs), from Stream Interaction Regions (SIRs) in the solar wind, from planetary magnetospheres and from galactic and anomalous cosmic rays. The lowest (suprathermal) energy part of the spectrum shows high variability and probably includes contributions from previous SEP events and from some quasi-continuous ion acceleration process operating in the solar atmosphere or in the interplanetary medium. The EPD instrument suite onboard Solar Orbiter is designed to measure all these energetic particle populations, from suprathermal energies up to the lowest energy part of the galactic and anomalous cosmic ray spectrum (affected by solar modulation).  

Solar Orbiter has been designed to provide an efficient combination of RS and IS observations. This means that EPD will contribute to achieve all the mission science goals, however its role will be especially relevant to address the third goal: ``How do solar eruptions produce energetic particle radiation that fills the heliosphere?''. This question can be broken down into several key topics:

\begin{itemize}
	\item What are the seed populations for energetic particles?
	\item How and where are energetic particles accelerated at the Sun?
	\item How are energetic particles released from their sources and distributed in space and time?
\end{itemize}

Suprathermal particles with energies above the ambient plasma in the outer corona and the solar wind are known to play an important role as seed population for acceleration during SEP events. The variability of this seed population may be a key factor to explain the wide range of intensities and composition observed in SEP events. EPD will have the opportunity to perform IS measurements of the composition and temporal variations of the suprathermal seed population close to the Sun, contributing also to understand the origins of the suprathermal ion pool itself. 

Energetic particles escaping from the acceleration sites continue their propagation through the turbulent interplanetary magnetic field. As shown by Helios observations \citep{Wibberenz_2006}, SEP events close to the Sun are much less disturbed by interplanetary transport effects compared to 1 AU observations. As the observing s/c goes farther from the Sun, the interplanetary scattering effects become more important and often multiple injections closely spaced in time cannot be resolved. For this reason, Solar Orbiter observations close to the perihelion will be crucial to unveil SEP injection, acceleration, trapping, release and transport processes. These observations will contribute to solve the controversy about the SEP acceleration sites, disentangling the contribution of acceleration at CME-driven shocks and at reconnection sites in solar flares or behind CMEs. 

EPD consists of four instruments measuring energetic electrons, protons and ions, operating at partly overlapping energy ranges covering from few keV to 450 MeV/nuc: 

\begin{itemize}
	\item SupraThermal Electrons and Protons (STEP)
	\item Suprathermal Ion Spectrograph (SIS)
	\item Electron Proton Telescope (EPT)
	\item High Energy Telescope (HET)
\end{itemize}

\begin{figure}
\includegraphics[width=70mm]{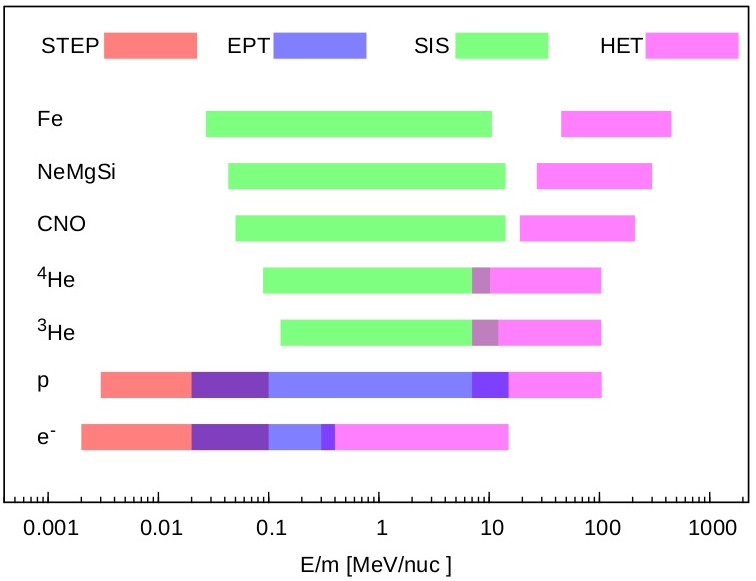}
\caption{Energy windows covered by the EPD instruments for different species.}
\label{figEPDwindows}
\end{figure}

\begin{figure}
\includegraphics[width=75mm]{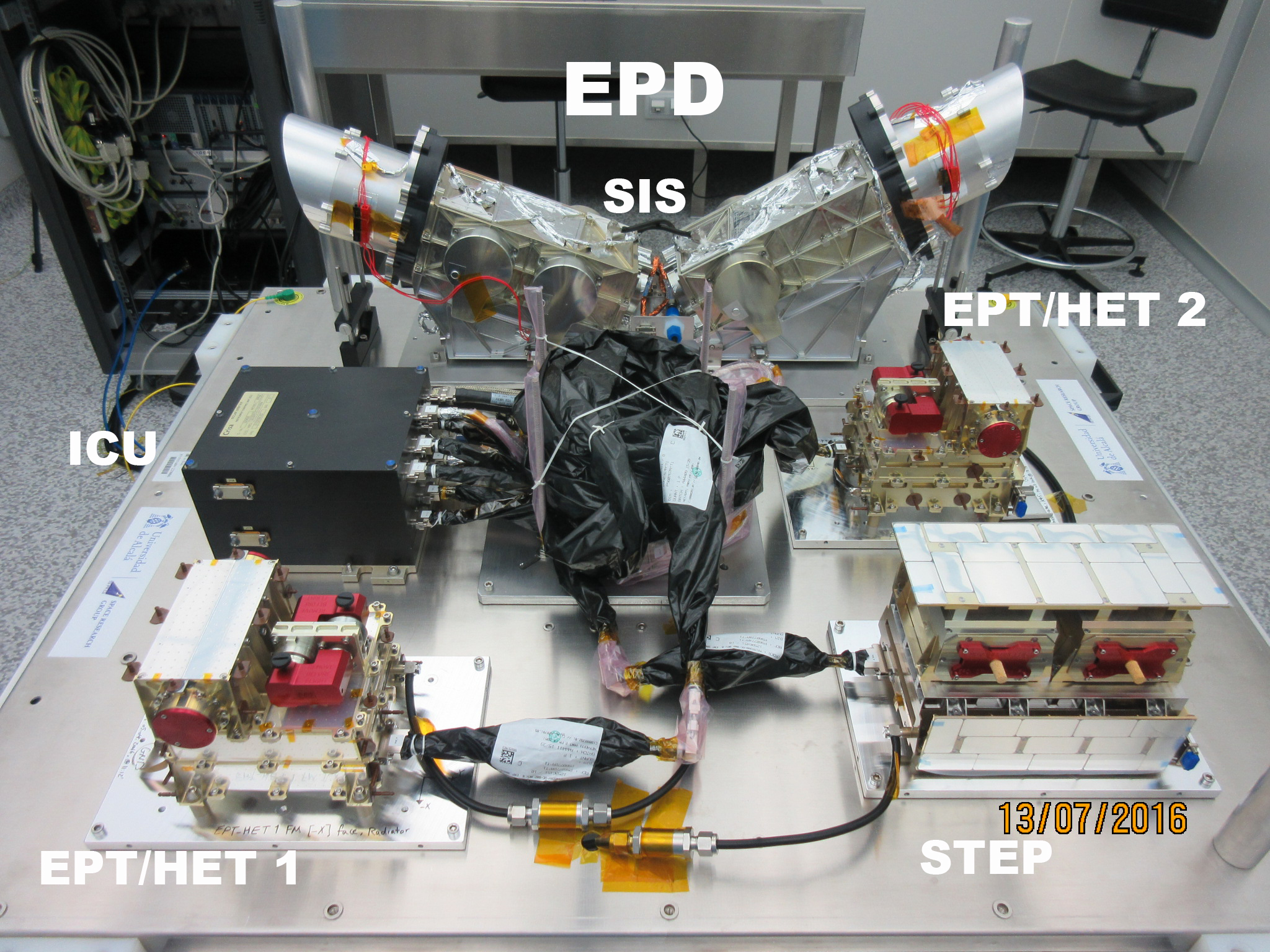}
\caption{EPD instrument suite during the integration tests performed in July 2016.}
\label{figEPDintegration}
\end{figure}

The energy intervals covered by the different instruments for various particle species are summarized in Figure \ref{figEPDwindows}. The four EPD sensors share a common Instrument Control Unit (ICU). EPD instruments and the ICU have significant heritage from previous missions, improved and optimized for the close approach to the Sun. Figure \ref{figEPDintegration} shows a picture of the whole EPD instrument suite during the integration tests performed in July 2016.

\begin{figure}
\includegraphics[width=75mm]{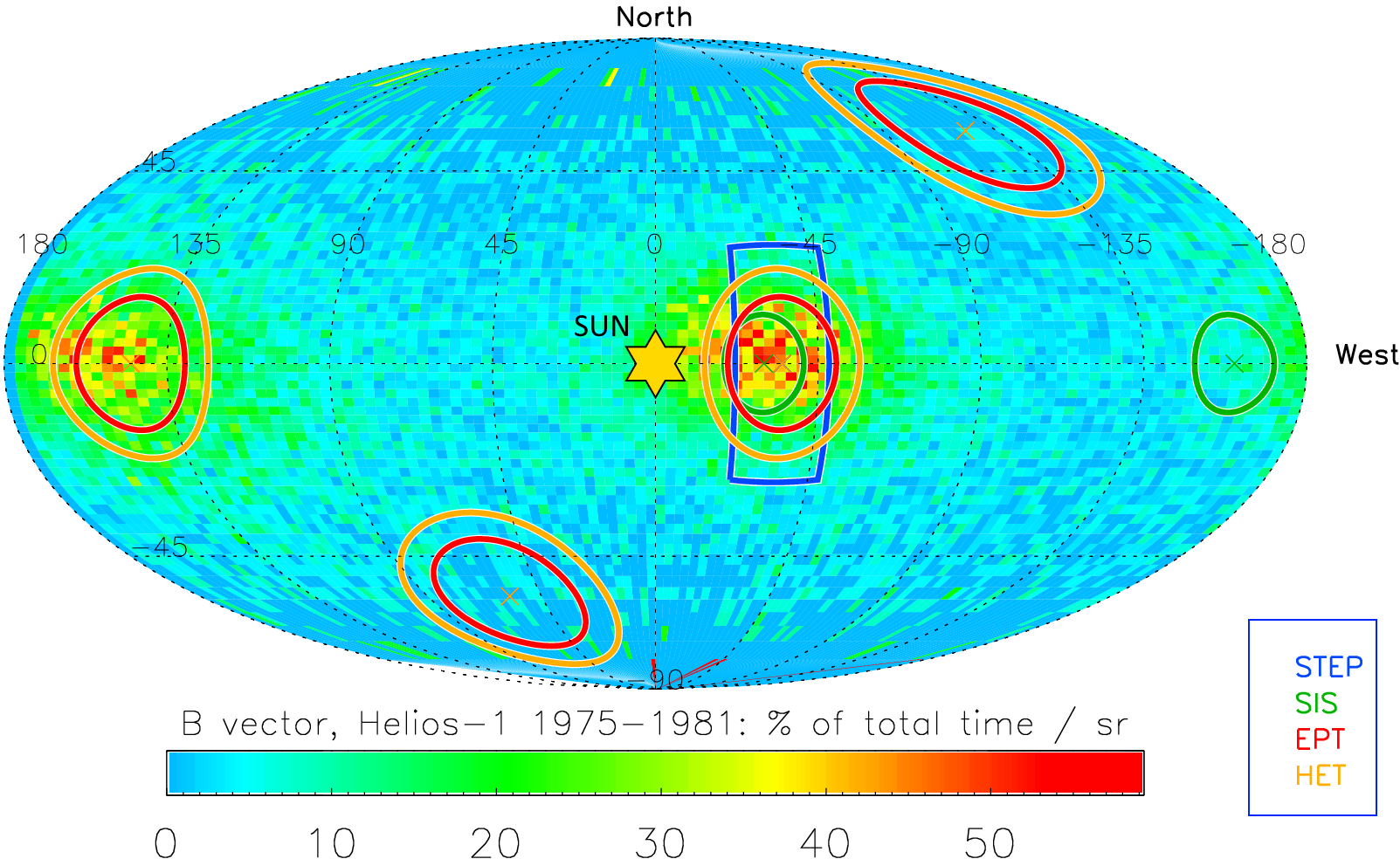}
\caption{EPD sensors fields of view.}
\label{figEPDfovs}
\end{figure}

The EPD sensors will measure the composition, spectra and anisotropies of energetic particles with sufficient temporal, spectral, angular and mass resolution to achieve the mission science goals. The geometric factors are scaled to avoid saturation by high particle fluxes close to the perihelia. Since Solar Orbiter is a three-axis stabilized s/c, EPD uses multiple apertures and sectoring to cover different pointing directions, providing information about the directional distribution of energetic particles reaching the s/c. This information combined with the magnetic field data will be used to obtain energetic particle pitch angle distributions, of fundamental importance to understand the interplanetary propagation of SEPs. Figure \ref{figEPDfovs} shows the fields of view of the different EPD sensors in the s/c reference frame. The background is color-coded as a function of the the interplanetary magnetic field vector distribution observed by the Helios mission. The different units which constitute EPD are described in the following sections.

\subsection{Suprathermal Electrons and Protons (STEP)}
\begin{figure}
\includegraphics[width=70mm]{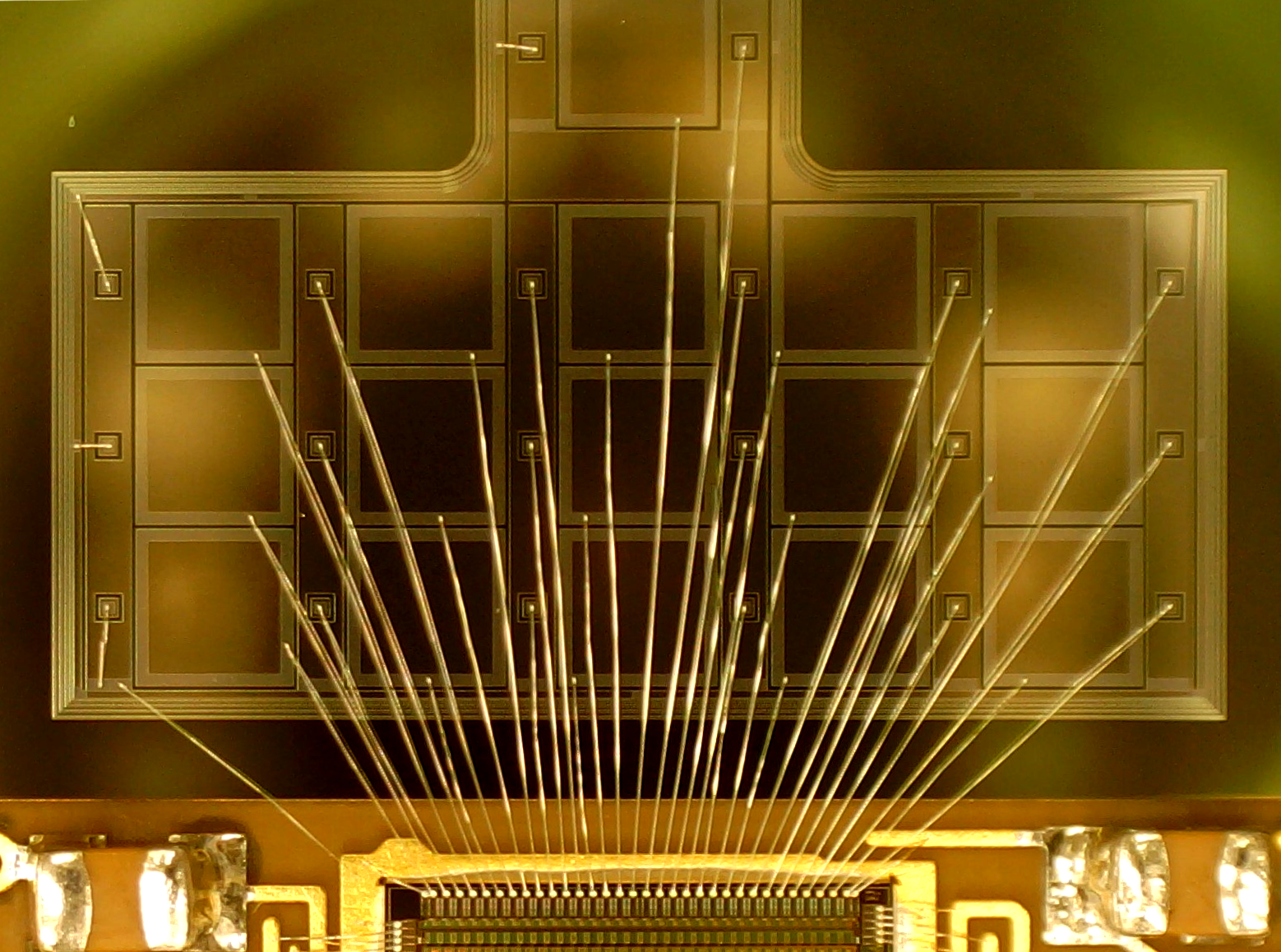}
\caption{STEP pixelated SSD. The top pixel dedicated to monitor the background produced by galactic cosmic ray and the bottom 3$\times$5 pixel array used to obtain directional information.}
\label{figSTEPpixels}
\end{figure}

\begin{figure}
\includegraphics[width=70mm]{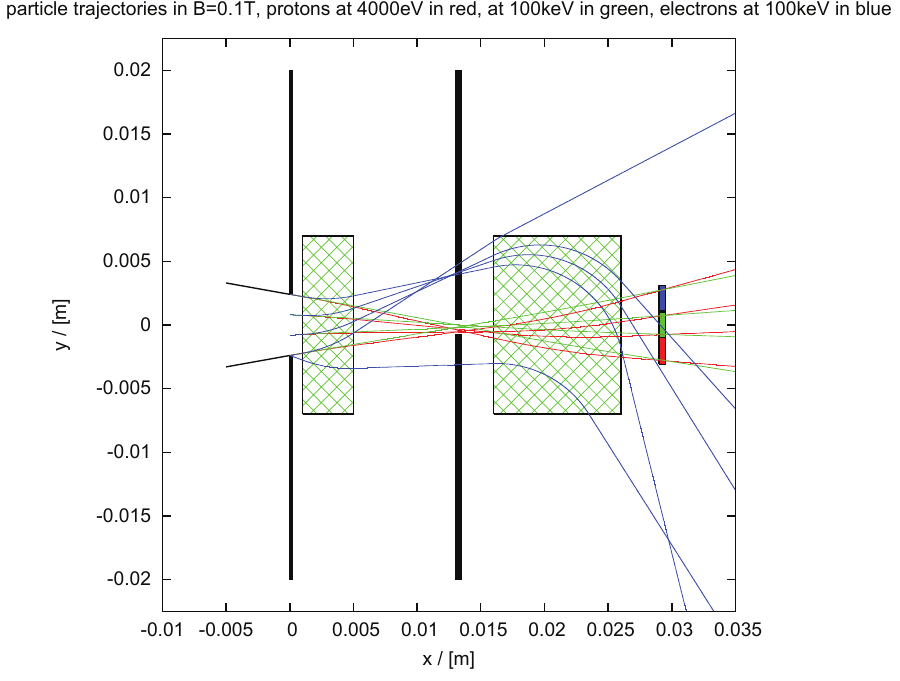}
\caption{Simulation of STEP magnetic deflection system. Two permanent magnets (green rectangles) effectively deflect electrons (blue trajectories) while protons within the instrument energy range (green and red trajectories) remain almost unaffected. The black vertical solid lines symbolize the entrance system of STEP (aperture and pinhole). Since the deflected electrons do not cross the pinhole, they would stop at the entrance system.}
\label{figSTEPmagnet}
\end{figure}

\begin{figure}
\includegraphics[width=70mm]{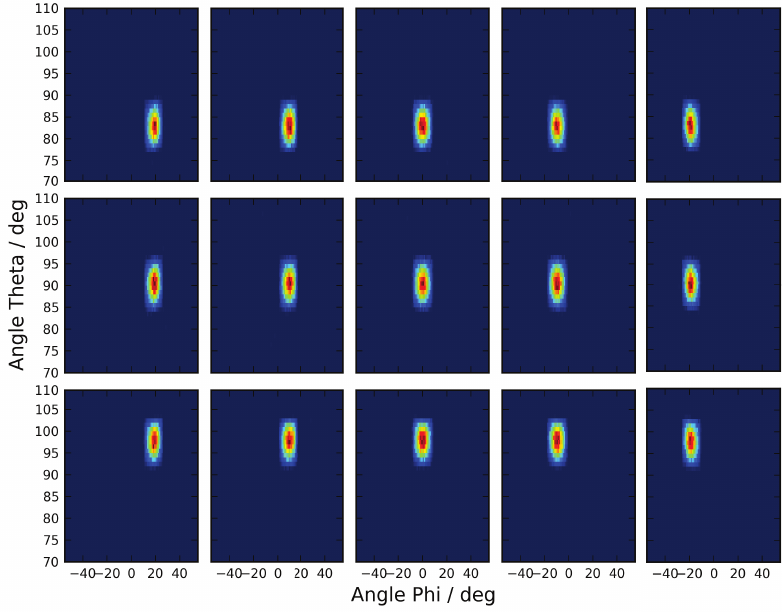}
\caption{Monte Carlo simulation results of the angular resolution achieved by the STEP 3$\times$5 pixelated proton detector (equipped with magnet).}
\label{figSTEPsimu}
\end{figure}

STEP will measure electrons and ions in the suprathermal energy range, covering 2-100 keV for electrons and 3-100 keV for protons. It has heritage from the Suprathermal Electron Instrument (STE, \cite{Lin_2008}) onboard the Solar Terrestrial Relations Observatory (STEREO), and consists of two identical units sharing common electronics. Both units have identical rectangular 54$^{\circ}\times$26$^{\circ}$ fields of view, pointing sunward along the nominal direction of the Parker spiral. Both apertures consists of a pinhole in combination with baffles at the entrance. This configuration reduces the amount of stray-light on the detector. Each unit has a single layer of silicon solid-state detectors (SSDs) segmented into 16 2$\times$2 mm$^2$ pixels (see Figure \ref{figSTEPpixels}). One of these pixels is used to monitor the background contribution from galactic cosmic rays, while the other 15 pixels combined with the pinhole aperture provide directional information. The use of SSDs with ultra-thin ohmic contacts provides high sensitivity compared to traditional electrostatic analyzers used for solar wind electron instruments. STEP is able to measure electron and ion fluxes with up to 1 s cadence. One of the unit's sensors is equipped with a magnetic deflection system which rejects electrons, while leaving ion trajectories almost unaffected (see Figure \ref{figSTEPmagnet}). This unit will provide ion fluxes while the second unit will measure both, electrons and ions. The difference between both measurements will be used to obtain the electron flux. The nominal geometric factor of each STEP unit is $7.5\cdot10^{-3}$ cm$^2$sr. During periods with very high fluxes the geometric factor can be reduced to $1.7\cdot10^{-4}$ cm$^2$sr by reducing the active area of the pixels to 0.3$\times$0.3 mm$^2$. Monte Carlo simulations results showing the angular resolution capabilities of the STEP proton telescope are shown in Figure \ref{figSTEPsimu}.

\subsection{Suprathermal Ion Spectrograph (SIS)}
\begin{figure}
\includegraphics[width=70mm]{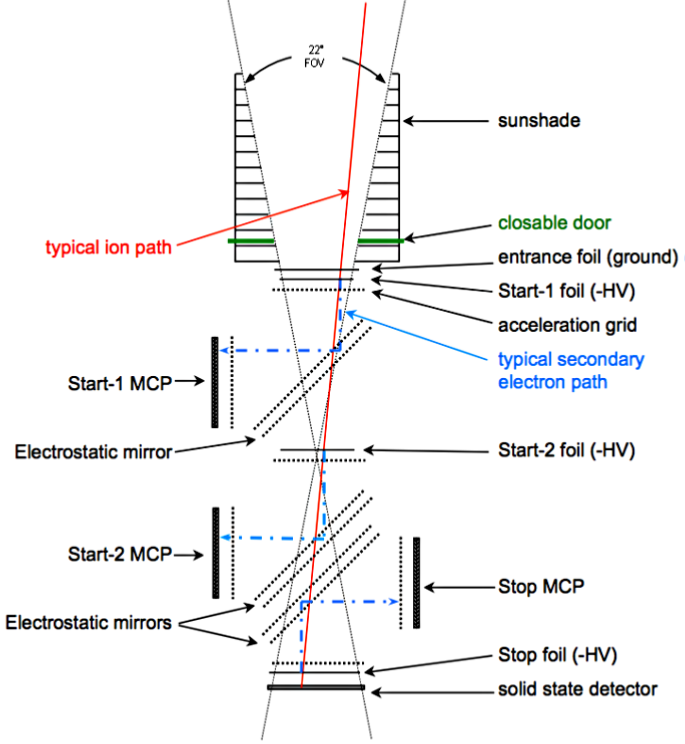}
\caption{Cross-section view of the SIS sensor.}
\label{figSISsketch}
\end{figure}

\begin{figure}
\includegraphics[width=40mm]{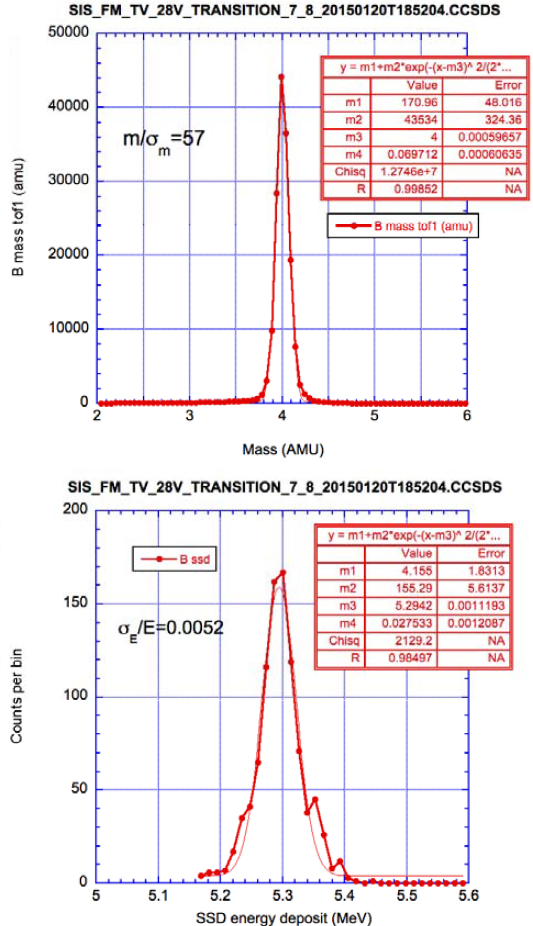}
\caption{SIS mass and energy resolution obtained using a $^{241}$Am radioactive source.}
\label{figSIScalib}
\end{figure}

SIS is a time-of-flight mass spectrometer that will measure all elements from He to Fe, sampling also trans-iron elements. The energy window is species-dependent, covering between 50 keV/nuc and 14 MeV/n for CNO. The instrument design has heritage from the Ultra-Low-Energy Isotope Spectrometer (ULEIS, \cite{Mason_1998}) onboard the Advanced Composition Explorer (ACE) and the Suprathermal Ion Telescope (SIT, \cite{Mason_2008a}) onboard STEREO. SIS consists of two particle telescopes, one looking sunward along the nominal Parker spiral direction and the other looking approximately in the anti-sunward direction, 130$^{\circ}$ away from the sunward-pointing telescope. Each telescope has a conical field of view with a full aperture of 22$^{\circ}$ and a nominal geometric factor of 0.2 cm$^2$sr. Both telescopes share a single electronics box. A sketch of one of the SIS telescopes is shown in Figure \ref{figSISsketch}. Time of flight information is collected when the ion passes through the \textit{Start-1}, \textit{Start-2}, and \textit{Stop} detector foils and secondary electrons are emitted, accelerated to $\sim$1 kV, and directed via isochronous mirrors onto microchannel plate stacks. Finally, they deposit their energy in the SSD at the back of the instrument. The combination of the total energy and time of fight information permits particle identification. SIS will acquire data with a nominal cadence of 30 s, being able to achieve 3 s during burst mode periods. The very high mass resolution of m$/\sigma_m\sim$ 50 will allow SIS to measure $^3$He$/^4$He ratios with uncertainty $<1\%$. Figure \ref{figSIScalib} shows the mass and energy resolution achieved by one of the SIS telescopes (proto flight model) obtained during the calibration tests performed using an $^{241}$Am alpha particle source. These calibration tests showed that SIS meets or exceeds the mission requirements. 

\subsection{Electron-Proton Telescope (EPT)}
\label{EPT}
\begin{figure}
\includegraphics[width=70mm]{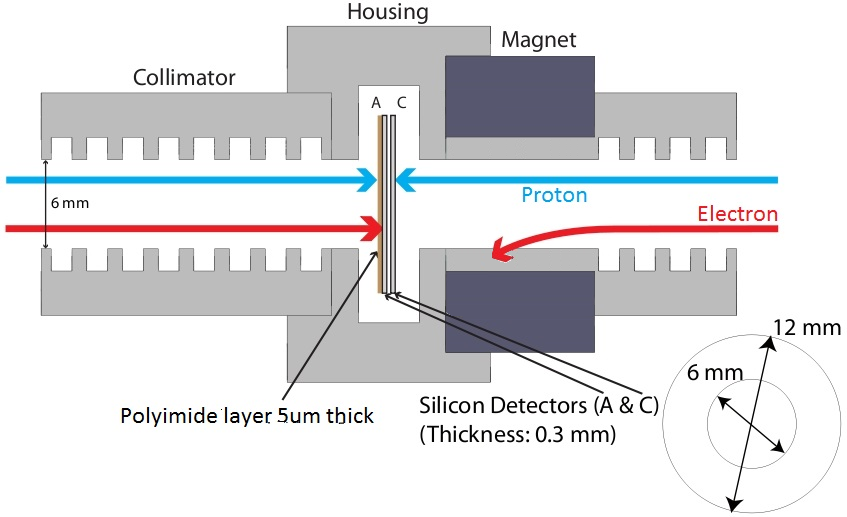}
\caption{Cross-section view of an EPT double-ended telescope.}
\label{figEPTsketch}
\end{figure}

\begin{figure}
\includegraphics[width=65mm]{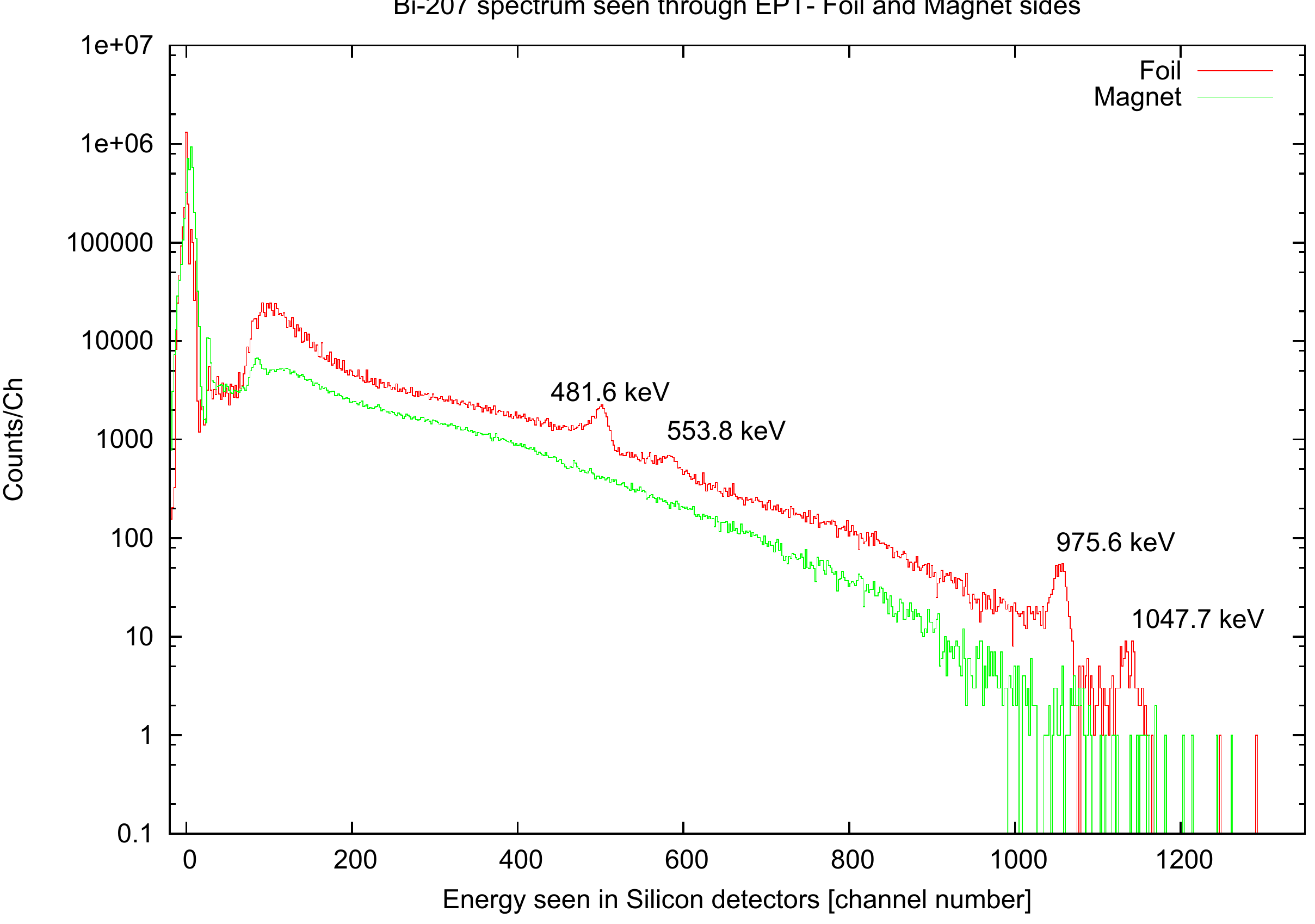}
\caption{$^{207}$Bi radioactive source spectra obtained by the EPT engineering model. The electron peaks disappear for the EPT aperture equipped with a magnet, meaning that the deflection system effectively rejects electrons.}
\label{figEPTsimu}
\end{figure}

EPT will measure 20-400 keV electrons and 20 keV-7 MeV protons. It has direct heritage from the Solar Electron and Proton Telescope (SEPT, \cite{Muller-Mellin_2008}) instrument onboard STEREO, using a magnet-foil technique for particle separation. Each double-ended EPT telescope has two closely spaced SSDs operating in anti-coincidence. As shown in Figure \ref{figEPTsketch}, one SSD looks through a polyimide foil while the second SSD looks through a magnetic deflection system. The foil stops low-energy protons while leaving electrons mostly unaffected. The magnet system effectively deflects electrons leaving protons unaffected. Since the magnets corresponding to each pair of double-ended telescopes are closely placed together and form a compensating pair of dipoles, the long-range field is strongly attenuated, minimizing the disturbance of the measurements by the magnetometer onboard Solar Orbiter. Figure \ref{figEPTsimu} shows EPT engineering model calibration results with a $^{207}$Bi radioactive source demonstrating the effective rejection of electrons by the magnet system. There are two EPT units, each one consisting of two double-ended telescopes. This setup provides a total of four view directions. The first EPT unit apertures point along the nominal Parker spiral in sunward and anti-sunward direction. The second unit apertures point 56$^{\circ}$ above and below the orbital plane. The electronics box of each EPT unit is shared with a HET unit. Each EPT aperture has a conical field of view with a full aperture of 30$^{\circ}$ and a nominal geometric factor of 0.01 cm$^2$sr. The maximum time cadence is 1 s.

\subsection{High Energy Telescope (HET)}
\begin{figure}
\includegraphics[width=60mm]{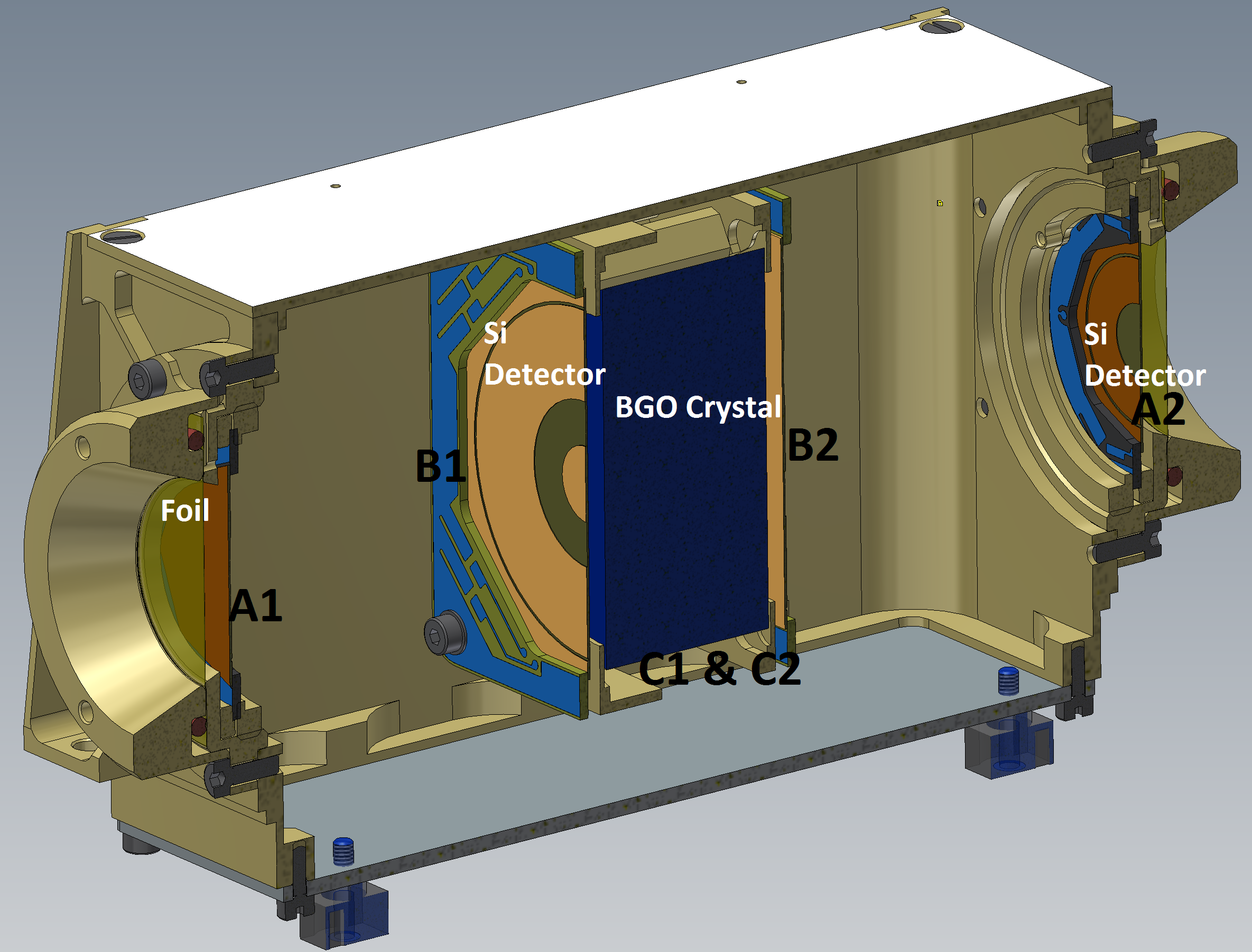}
\caption{Cross-section view of a HET double-ended telescope. Note the large BGO scintillator located in the center and the two SSDs located at both sides.}
\label{figHETsketch}
\end{figure}

\begin{figure}
\includegraphics[width=70mm]{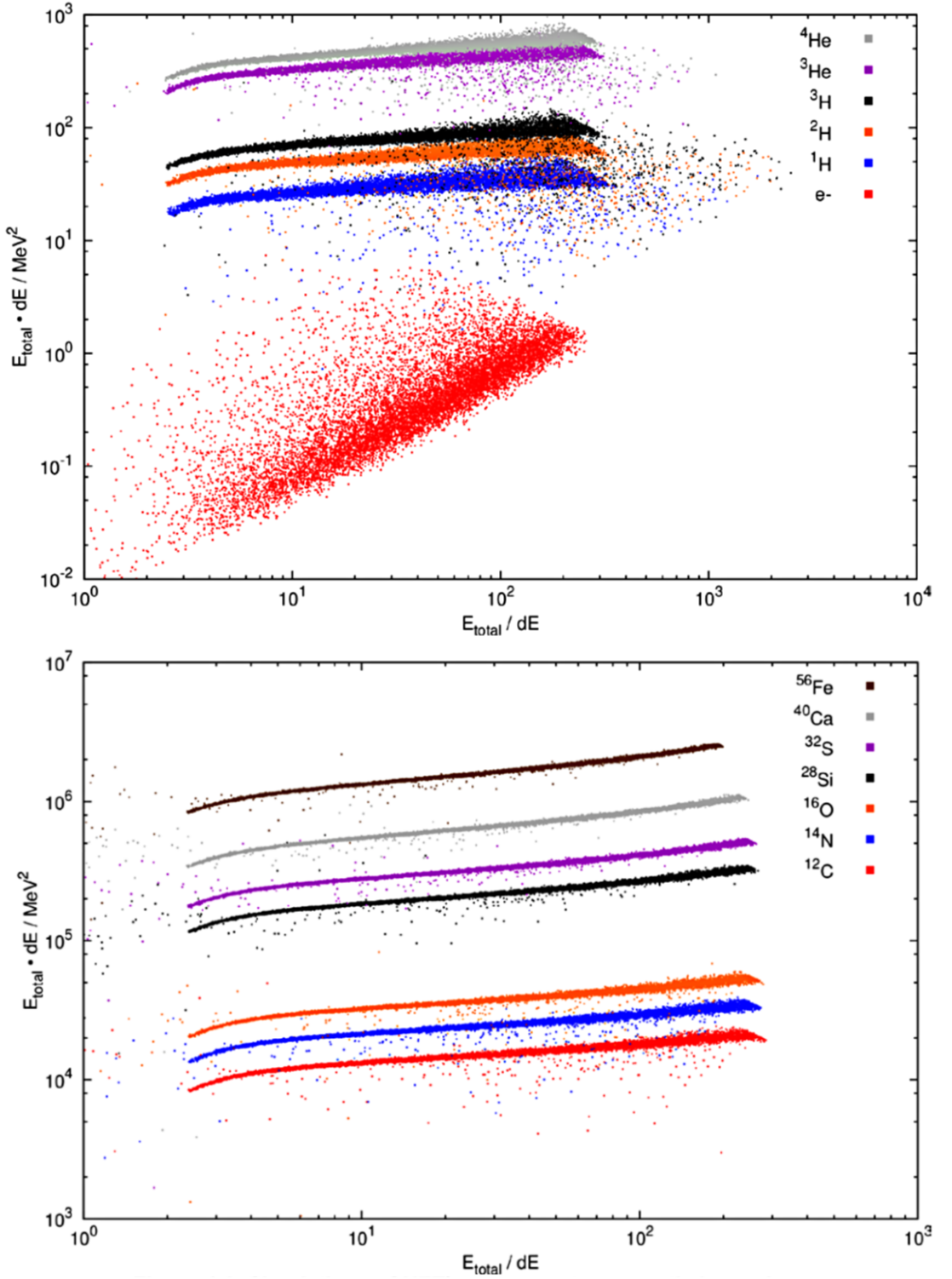}
\caption{Monte Carlo simulation results of HET response to light ions and electrons (top) and heavy ions (bottom). Each plot shows the product of the energy loss in the SSD and the total energy versus their ratio.}
\label{figHETsimu}
\end{figure}

HET covers the energy range, which is of specific interest for space weather, and will perform the measurements needed to understand the origin of high-energy SEP events at the Sun. HET will measure electrons between 300 keV and 15 MeV, protons from 10 to 100 MeV and ions from 20 MeV/nuc to 450 MeV/nuc (the exact interval is Z-dependent, see \cite{Tammen_2015}). Incident particles will be identified using the dE/dx vs. total energy technique. It has heritage from the Radiation Assessment Detector (RAD, \cite{Hassler_2012}) onboard the Mars Science Laboratory (MSL). HET consists of two double-ended sensor heads, one pointing sunward and anti-sunward along the nominal Parker spiral, the other pointing above and below the orbital plane. Thus, HET has a total of four viewing directions (analogous to EPT, see section \ref{EPT} and Figure \ref{figEPDfovs}). Both HET sensors are identical and consist of a double-ended set of SSDs and a a high-density Bismuth Germanate (BGO) calorimeter scintillator (Figure \ref{figHETsketch}). Each double HET sensor unit shares the electronic box with an EPT unit. HET allows separation of the helium isotopes down to a $^3$He$/^4$He isotope ratio of about 1$\%$ in a limited energy range. Figure \ref{figHETsimu} show Monte Carlo simulation results illustrating HET response to heavy ions, light ions and electrons. HET has a conical field of view with a full aperture of 43$^{\circ}$ and a nominal geometric factor of 0.27 cm$^2$sr. The front detectors of HET (both sides) are protected by laminated Kapton+Al foil (Kapton is 50 ${\mu}$m and Al is 25 ${\mu}$m thick) to reduce low energy particle flux. In addition, the front detector is divided into concentric segments that allow the reduction of proton count rates during high intensity events. In such situations, the thresholds on the larger segment are increased to beyond the energy deposit of protons. This scheme retains the detection power for the much rarer heavy ions while reducing the counting rate for the abundant protons. The maximum instrument cadence is 1 s.

\subsection{Instrument Control Unit (ICU)}
The ICU provides a single point of connection between the s/c and all the EPD sensors, acting as data and power interface. It is composed of the Common Data Processing Unit (CDPU) and the Low Voltage Power Supply (LVPS). The ICU shares information with other EPD instruments to allow synchronized burst-mode operations following on-board identification of predefined triggering events in the EPD data. The ICU has strong architecture heritage from the Common Data Processing Unit (CDPU) for the Comprehensive Suprathermal and Energetic Particle Analyzer (COSTEP, \cite{Muller-Mellin_1995}) and the Energetic and Relativistic Nuclei and Electron (ERNE, \cite{Torsti_1995}) instruments onboard the Solar and Heliospheric Observatory (SOHO). The ICU is designed to manage sensor's control and monitoring, timing clock, and data collection, compression, and packetization for telemetry. ICU is also responsible for the s/c telecommand reception and delivery to the sensors if necessary. The CDPU in based on a LEON2 soft-processor implemented in a RTAX2000 FPGA from Actel, it contains external RAM, EEPROM and PROM memories, two hot redundant SpaceWire interfaces, and four identical serial links (UART-LVDS) with the sensors. The CDPU PROM contains the boot code and it can be used to load flight code from either the EEPROM, that contains two copies of the application software, or from the s/c interface via telecommand. The LVPS board is responsible for filtering, monitoring and switching the s/c primary power. It also provides the power supply to both CDPUs. 

\subsection{EPD data products}
EPD has a total telemetry budget of 3600 bit/s (housekeeping $+$ science data). Solar Orbiter telemetry rate is highly variable during the orbit. This implies that data will be downliked to ground with strongly varying latencies of up to several months. A minimal set of science data from all the instruments will be downlinked daily (low latency dataset), mainly for planning/monitoring  purposes, but also with scientific value. In order to optimize the science return fulfilling telemetry limitations, EPD science data sent to ground will vary depending on radial distance, having higher cadences close to the perihelia. 

Following in-orbit commissioning, the Principal Investigators of the different instruments onboard Solar Orbiter retain exclusive data rights for the purpose of calibration and verification for a period of 3 months after the receipt of the original science telemetry. Upon delivery of data to the ESA Science Operations Center, they will be made available to the scientific community through the ESA science data archive. The instrument teams will provide records of processed data with all relevant information on calibration and instrument properties to the ESA science data archive, which will be the repository of all mission products \citep{Marsden_2012}. Solar Orbiter's IS instruments will provide processed data using the Common Data Format (CDF) standard.

\section{MULTIPOINT OBSERVATION OPPORTUNITIES}
Solar Orbiter will offer excellent opportunities for multi-point observation campaigns combining measurements by multiple s/c. Synergies with the SPP mission are of particular relevance, since both missions have overlapping timelines and the SPP perihelion, reaching up to $<$10 solar radii, will permit IS observations at the SEP acceleration region close to the Sun simultaneous to IS measurements at larger radial distances and with continuous RS coverage provided by Solar Orbiter and near-Earth s/c. Certain geometric configurations will be particularly appropriate for multi-point measurements in order to optimize the science return:
\begin{itemize}
	\item Close approaches of Solar Orbiter to SPP or other s/c providing the opportunity for cross-calibration of the particle instruments onboard.
	\item Radial alignments enabling the observation of plasma ``packets'' from the same solar source region at progressive radial distances as well the study of energetic particle radial gradients.
	\item Alignments along the same interplanetary magnetic field line allowing the observation of SEPs originating at the same acceleration site by two or more s/c located at different radial distances.
	\item Combination of RS observations of near-limb source regions and IS plasma observations by s/c with angular separations close to 90$^{\circ}$.
	\item Observations of SEP events by multiple s/c covering wide angular regions (both, in longitude and latitude) in order to investigate the spatial distribution of SEPs and the physical mechanisms producing wide-spread SEPs events (see e.g. \cite{Lario_2013, Dresing_2014b}). These observations will help to understand the possible role played by interplanetary cross-field diffusion, acceleration at wide shocks and distorted coronal magnetic fields with large latitudinal/longitudinal deviations from the radial direction.
\end{itemize}

\section {SUMMARY AND CONCLUSIONS}
Solar Orbiter is a unique mission conceived to unveil the Sun-heliosphere connection.  The orbital configuration includes a close perihelion, high inclination intervals allowing the observation of the solar polar regions and quasi-co-rotation periods. These orbital characteristics are combined with a comprehensive combination of IS and RS instruments and excellent opportunities for multi-s/c studies. The EPD suite will provide high-quality energetic particle observations over a wide energy range and multiple species, key to understand the seed populations, injection, acceleration and transport processes of SEPs.

\bigskip 
\begin{acknowledgments}
The authors acknowledge the financial support of the Spanish MINECO under projects ESP2013-48346-C2-1-R and ESP2015-68266-R (MINECO/FEDER). SIS was funded by ESA as a European-led facility instrument. CAU acknowledges the financial support from the German space Agency (DLR) under grants 50OT1002 and 50OT1202 and by the University of Kiel.
\end{acknowledgments}

\bigskip 
\bibliography{rghbib}

\begin{thebibliography}{16}
\expandafter\ifx\csname natexlab\endcsname\relax\def\natexlab#1{#1}\fi
\expandafter\ifx\csname bibnamefont\endcsname\relax
  \def\bibnamefont#1{#1}\fi
\expandafter\ifx\csname bibfnamefont\endcsname\relax
  \def\bibfnamefont#1{#1}\fi
\expandafter\ifx\csname citenamefont\endcsname\relax
  \def\citenamefont#1{#1}\fi
\expandafter\ifx\csname url\endcsname\relax
  \def\url#1{\texttt{#1}}\fi
\expandafter\ifx\csname urlprefix\endcsname\relax\def\urlprefix{URL }\fi
\providecommand{\bibinfo}[2]{#2}
\providecommand{\eprint}[2][]{\url{#2}}

\bibitem[{\citenamefont{{M{\"u}ller} et~al.}(2013)\citenamefont{{M{\"u}ller},
  {Marsden}, {St.~Cyr}, and {Gilbert}}}]{Muller_2013}
\bibinfo{author}{\bibfnamefont{D.}~\bibnamefont{{M{\"u}ller}}},
  \bibinfo{author}{\bibfnamefont{R.~G.} \bibnamefont{{Marsden}}},
  \bibinfo{author}{\bibfnamefont{O.~C.} \bibnamefont{{St.~Cyr}}},
  \bibnamefont{and} \bibinfo{author}{\bibfnamefont{H.~R.}
  \bibnamefont{{Gilbert}}}, \bibinfo{journal}{Solar Physics}
  \textbf{\bibinfo{volume}{285}}, \bibinfo{pages}{25} (\bibinfo{year}{2013}),
  \eprint{1207.4579}.

\bibitem[{\citenamefont{{Fox} et~al.}(2016)\citenamefont{{Fox}, {Velli},
  {Bale}, {Decker}, {Driesman}, {Howard}, {Kasper}, {Kinnison}, {Kusterer},
  {Lario} et~al.}}]{Fox_2016}
\bibinfo{author}{\bibfnamefont{N.~J.} \bibnamefont{{Fox}}},
  \bibinfo{author}{\bibfnamefont{M.~C.} \bibnamefont{{Velli}}},
  \bibinfo{author}{\bibfnamefont{S.~D.} \bibnamefont{{Bale}}},
  \bibinfo{author}{\bibfnamefont{R.}~\bibnamefont{{Decker}}},
  \bibinfo{author}{\bibfnamefont{A.}~\bibnamefont{{Driesman}}},
  \bibinfo{author}{\bibfnamefont{R.~A.} \bibnamefont{{Howard}}},
  \bibinfo{author}{\bibfnamefont{J.~C.} \bibnamefont{{Kasper}}},
  \bibinfo{author}{\bibfnamefont{J.}~\bibnamefont{{Kinnison}}},
  \bibinfo{author}{\bibfnamefont{M.}~\bibnamefont{{Kusterer}}},
  \bibinfo{author}{\bibfnamefont{D.}~\bibnamefont{{Lario}}},
  \bibnamefont{et~al.}, \bibinfo{journal}{Space Science Reviews}
  \textbf{\bibinfo{volume}{204}}, \bibinfo{pages}{7} (\bibinfo{year}{2016}).

\bibitem[{\citenamefont{{Porsche}}(1977)}]{Porsche_1977}
\bibinfo{author}{\bibfnamefont{H.}~\bibnamefont{{Porsche}}},
  \bibinfo{journal}{Journal of Geophysics Zeitschrift Geophysik}
  \textbf{\bibinfo{volume}{42}}, \bibinfo{pages}{551} (\bibinfo{year}{1977}).

\bibitem[{\citenamefont{{Benkhoff} et~al.}(2010)\citenamefont{{Benkhoff}, {van
  Casteren}, {Hayakawa}, {Fujimoto}, {Laakso}, {Novara}, {Ferri}, {Middleton},
  and {Ziethe}}}]{Benkhoff_2010}
\bibinfo{author}{\bibfnamefont{J.}~\bibnamefont{{Benkhoff}}},
  \bibinfo{author}{\bibfnamefont{J.}~\bibnamefont{{van Casteren}}},
  \bibinfo{author}{\bibfnamefont{H.}~\bibnamefont{{Hayakawa}}},
  \bibinfo{author}{\bibfnamefont{M.}~\bibnamefont{{Fujimoto}}},
  \bibinfo{author}{\bibfnamefont{H.}~\bibnamefont{{Laakso}}},
  \bibinfo{author}{\bibfnamefont{M.}~\bibnamefont{{Novara}}},
  \bibinfo{author}{\bibfnamefont{P.}~\bibnamefont{{Ferri}}},
  \bibinfo{author}{\bibfnamefont{H.~R.} \bibnamefont{{Middleton}}},
  \bibnamefont{and} \bibinfo{author}{\bibfnamefont{R.}~\bibnamefont{{Ziethe}}},
  \bibinfo{journal}{Planetary and Space Science} \textbf{\bibinfo{volume}{58}},
  \bibinfo{pages}{2} (\bibinfo{year}{2010}).

\bibitem[{\citenamefont{{Wibberenz} and {Cane}}(2006)}]{Wibberenz_2006}
\bibinfo{author}{\bibfnamefont{G.}~\bibnamefont{{Wibberenz}}} \bibnamefont{and}
  \bibinfo{author}{\bibfnamefont{H.~V.} \bibnamefont{{Cane}}},
  \bibinfo{journal}{\apj} \textbf{\bibinfo{volume}{650}}, \bibinfo{pages}{1199}
  (\bibinfo{year}{2006}).

\bibitem[{\citenamefont{{Lin} et~al.}(2008)\citenamefont{{Lin}, {Curtis},
  {Larson}, {Luhmann}, {McBride}, {Maier}, {Moreau}, {Tindall}, {Turin}, and
  {Wang}}}]{Lin_2008}
\bibinfo{author}{\bibfnamefont{R.~P.} \bibnamefont{{Lin}}},
  \bibinfo{author}{\bibfnamefont{D.~W.} \bibnamefont{{Curtis}}},
  \bibinfo{author}{\bibfnamefont{D.~E.} \bibnamefont{{Larson}}},
  \bibinfo{author}{\bibfnamefont{J.~G.} \bibnamefont{{Luhmann}}},
  \bibinfo{author}{\bibfnamefont{S.~E.} \bibnamefont{{McBride}}},
  \bibinfo{author}{\bibfnamefont{M.~R.} \bibnamefont{{Maier}}},
  \bibinfo{author}{\bibfnamefont{T.}~\bibnamefont{{Moreau}}},
  \bibinfo{author}{\bibfnamefont{C.~S.} \bibnamefont{{Tindall}}},
  \bibinfo{author}{\bibfnamefont{P.}~\bibnamefont{{Turin}}}, \bibnamefont{and}
  \bibinfo{author}{\bibfnamefont{L.}~\bibnamefont{{Wang}}},
  \bibinfo{journal}{Space Science Reviews} \textbf{\bibinfo{volume}{136}},
  \bibinfo{pages}{241} (\bibinfo{year}{2008}).

\bibitem[{\citenamefont{{Mason} et~al.}(1998)\citenamefont{{Mason}, {Gold},
  {Krimigis}, {Mazur}, {Andrews}, {Daley}, {Dwyer}, {Heuerman}, {James},
  {Kennedy} et~al.}}]{Mason_1998}
\bibinfo{author}{\bibfnamefont{G.~M.} \bibnamefont{{Mason}}},
  \bibinfo{author}{\bibfnamefont{R.~E.} \bibnamefont{{Gold}}},
  \bibinfo{author}{\bibfnamefont{S.~M.} \bibnamefont{{Krimigis}}},
  \bibinfo{author}{\bibfnamefont{J.~E.} \bibnamefont{{Mazur}}},
  \bibinfo{author}{\bibfnamefont{G.~B.} \bibnamefont{{Andrews}}},
  \bibinfo{author}{\bibfnamefont{K.~A.} \bibnamefont{{Daley}}},
  \bibinfo{author}{\bibfnamefont{J.~R.} \bibnamefont{{Dwyer}}},
  \bibinfo{author}{\bibfnamefont{K.~F.} \bibnamefont{{Heuerman}}},
  \bibinfo{author}{\bibfnamefont{T.~L.} \bibnamefont{{James}}},
  \bibinfo{author}{\bibfnamefont{M.~J.} \bibnamefont{{Kennedy}}},
  \bibnamefont{et~al.}, \bibinfo{journal}{Space Science Reviews}
  \textbf{\bibinfo{volume}{86}}, \bibinfo{pages}{409} (\bibinfo{year}{1998}).

\bibitem[{\citenamefont{{Mason} et~al.}(2008)\citenamefont{{Mason}, {Korth},
  {Walpole}, {Desai}, {von Rosenvinge}, and {Shuman}}}]{Mason_2008a}
\bibinfo{author}{\bibfnamefont{G.~M.} \bibnamefont{{Mason}}},
  \bibinfo{author}{\bibfnamefont{A.}~\bibnamefont{{Korth}}},
  \bibinfo{author}{\bibfnamefont{P.~H.} \bibnamefont{{Walpole}}},
  \bibinfo{author}{\bibfnamefont{M.~I.} \bibnamefont{{Desai}}},
  \bibinfo{author}{\bibfnamefont{T.~T.} \bibnamefont{{von Rosenvinge}}},
  \bibnamefont{and} \bibinfo{author}{\bibfnamefont{S.~A.}
  \bibnamefont{{Shuman}}}, \bibinfo{journal}{Space Science Reviews}
  \textbf{\bibinfo{volume}{136}}, \bibinfo{pages}{257} (\bibinfo{year}{2008}).

\bibitem[{\citenamefont{{M{\"u}ller-Mellin}
  et~al.}(2008)\citenamefont{{M{\"u}ller-Mellin}, {B{\"o}ttcher}, {Falenski},
  {Rode}, {Duvet}, {Sanderson}, {Butler}, {Johlander}, and
  {Smit}}}]{Muller-Mellin_2008}
\bibinfo{author}{\bibfnamefont{R.}~\bibnamefont{{M{\"u}ller-Mellin}}},
  \bibinfo{author}{\bibfnamefont{S.}~\bibnamefont{{B{\"o}ttcher}}},
  \bibinfo{author}{\bibfnamefont{J.}~\bibnamefont{{Falenski}}},
  \bibinfo{author}{\bibfnamefont{E.}~\bibnamefont{{Rode}}},
  \bibinfo{author}{\bibfnamefont{L.}~\bibnamefont{{Duvet}}},
  \bibinfo{author}{\bibfnamefont{T.}~\bibnamefont{{Sanderson}}},
  \bibinfo{author}{\bibfnamefont{B.}~\bibnamefont{{Butler}}},
  \bibinfo{author}{\bibfnamefont{B.}~\bibnamefont{{Johlander}}},
  \bibnamefont{and} \bibinfo{author}{\bibfnamefont{H.}~\bibnamefont{{Smit}}},
  \bibinfo{journal}{Space Science Reviews} \textbf{\bibinfo{volume}{136}},
  \bibinfo{pages}{363} (\bibinfo{year}{2008}).

\bibitem[{\citenamefont{{Tammen} et~al.}(2015)\citenamefont{{Tammen},
  {Elftmann}, {Kulkarni}, {B{\"o}ttcher}, and
  {Wimmer-Schweingruber}}}]{Tammen_2015}
\bibinfo{author}{\bibfnamefont{J.}~\bibnamefont{{Tammen}}},
  \bibinfo{author}{\bibfnamefont{R.}~\bibnamefont{{Elftmann}}},
  \bibinfo{author}{\bibfnamefont{S.~R.} \bibnamefont{{Kulkarni}}},
  \bibinfo{author}{\bibfnamefont{S.~I.} \bibnamefont{{B{\"o}ttcher}}},
  \bibnamefont{and} \bibinfo{author}{\bibfnamefont{R.~F.}
  \bibnamefont{{Wimmer-Schweingruber}}}, \bibinfo{journal}{Nuclear Instruments
  and Methods in Physics Research B} \textbf{\bibinfo{volume}{360}},
  \bibinfo{pages}{129} (\bibinfo{year}{2015}).

\bibitem[{\citenamefont{{Hassler} et~al.}(2012)\citenamefont{{Hassler},
  {Zeitlin}, {Wimmer-Schweingruber}, {B{\"o}ttcher}, {Martin}, {Andrews},
  {B{\"o}hm}, {Brinza}, {Bullock}, {Burmeister} et~al.}}]{Hassler_2012}
\bibinfo{author}{\bibfnamefont{D.~M.} \bibnamefont{{Hassler}}},
  \bibinfo{author}{\bibfnamefont{C.}~\bibnamefont{{Zeitlin}}},
  \bibinfo{author}{\bibfnamefont{R.~F.} \bibnamefont{{Wimmer-Schweingruber}}},
  \bibinfo{author}{\bibfnamefont{S.}~\bibnamefont{{B{\"o}ttcher}}},
  \bibinfo{author}{\bibfnamefont{C.}~\bibnamefont{{Martin}}},
  \bibinfo{author}{\bibfnamefont{J.}~\bibnamefont{{Andrews}}},
  \bibinfo{author}{\bibfnamefont{E.}~\bibnamefont{{B{\"o}hm}}},
  \bibinfo{author}{\bibfnamefont{D.~E.} \bibnamefont{{Brinza}}},
  \bibinfo{author}{\bibfnamefont{M.~A.} \bibnamefont{{Bullock}}},
  \bibinfo{author}{\bibfnamefont{S.}~\bibnamefont{{Burmeister}}},
  \bibnamefont{et~al.}, \bibinfo{journal}{Space Science Reviews}
  \textbf{\bibinfo{volume}{170}}, \bibinfo{pages}{503} (\bibinfo{year}{2012}).

\bibitem[{\citenamefont{{M{\"u}ller-Mellin}
  et~al.}(1995)\citenamefont{{M{\"u}ller-Mellin}, {Kunow}, {Flei{\ss}ner},
  {Pehlke}, {Rode}, {R{\"o}schmann}, {Scharmberg}, {Sierks}, {Rusznyak},
  {McKenna-Lawlor} et~al.}}]{Muller-Mellin_1995}
\bibinfo{author}{\bibfnamefont{R.}~\bibnamefont{{M{\"u}ller-Mellin}}},
  \bibinfo{author}{\bibfnamefont{H.}~\bibnamefont{{Kunow}}},
  \bibinfo{author}{\bibfnamefont{V.}~\bibnamefont{{Flei{\ss}ner}}},
  \bibinfo{author}{\bibfnamefont{E.}~\bibnamefont{{Pehlke}}},
  \bibinfo{author}{\bibfnamefont{E.}~\bibnamefont{{Rode}}},
  \bibinfo{author}{\bibfnamefont{N.}~\bibnamefont{{R{\"o}schmann}}},
  \bibinfo{author}{\bibfnamefont{C.}~\bibnamefont{{Scharmberg}}},
  \bibinfo{author}{\bibfnamefont{H.}~\bibnamefont{{Sierks}}},
  \bibinfo{author}{\bibfnamefont{P.}~\bibnamefont{{Rusznyak}}},
  \bibinfo{author}{\bibfnamefont{S.}~\bibnamefont{{McKenna-Lawlor}}},
  \bibnamefont{et~al.}, \bibinfo{journal}{Solar Physics}
  \textbf{\bibinfo{volume}{162}}, \bibinfo{pages}{483} (\bibinfo{year}{1995}).

\bibitem[{\citenamefont{{Torsti} et~al.}(1995)\citenamefont{{Torsti},
  {Valtonen}, {Lumme}, {Peltonen}, {Eronen}, {Louhola}, {Riihonen}, {Schultz},
  {Teittinen}, {Ahola} et~al.}}]{Torsti_1995}
\bibinfo{author}{\bibfnamefont{J.}~\bibnamefont{{Torsti}}},
  \bibinfo{author}{\bibfnamefont{E.}~\bibnamefont{{Valtonen}}},
  \bibinfo{author}{\bibfnamefont{M.}~\bibnamefont{{Lumme}}},
  \bibinfo{author}{\bibfnamefont{P.}~\bibnamefont{{Peltonen}}},
  \bibinfo{author}{\bibfnamefont{T.}~\bibnamefont{{Eronen}}},
  \bibinfo{author}{\bibfnamefont{M.}~\bibnamefont{{Louhola}}},
  \bibinfo{author}{\bibfnamefont{E.}~\bibnamefont{{Riihonen}}},
  \bibinfo{author}{\bibfnamefont{G.}~\bibnamefont{{Schultz}}},
  \bibinfo{author}{\bibfnamefont{M.}~\bibnamefont{{Teittinen}}},
  \bibinfo{author}{\bibfnamefont{K.}~\bibnamefont{{Ahola}}},
  \bibnamefont{et~al.}, \bibinfo{journal}{Solar Physics}
  \textbf{\bibinfo{volume}{162}}, \bibinfo{pages}{505} (\bibinfo{year}{1995}).

\bibitem[{\citenamefont{{Marsden}}(2012)}]{Marsden_2012}
\bibinfo{author}{\bibfnamefont{R.}~\bibnamefont{{Marsden}}},
  \bibinfo{type}{Tech. Rep.}, \bibinfo{institution}{European Space Agency
  (SOL-EST-PL-00880)} (\bibinfo{year}{2012}).

\bibitem[{\citenamefont{{Lario} et~al.}(2013)\citenamefont{{Lario}, {Aran},
  {G{\'o}mez-Herrero}, {Dresing}, {Heber}, {Ho}, {Decker}, and
  {Roelof}}}]{Lario_2013}
\bibinfo{author}{\bibfnamefont{D.}~\bibnamefont{{Lario}}},
  \bibinfo{author}{\bibfnamefont{A.}~\bibnamefont{{Aran}}},
  \bibinfo{author}{\bibfnamefont{R.}~\bibnamefont{{G{\'o}mez-Herrero}}},
  \bibinfo{author}{\bibfnamefont{N.}~\bibnamefont{{Dresing}}},
  \bibinfo{author}{\bibfnamefont{B.}~\bibnamefont{{Heber}}},
  \bibinfo{author}{\bibfnamefont{G.~C.} \bibnamefont{{Ho}}},
  \bibinfo{author}{\bibfnamefont{R.~B.} \bibnamefont{{Decker}}},
  \bibnamefont{and} \bibinfo{author}{\bibfnamefont{E.~C.}
  \bibnamefont{{Roelof}}}, \bibinfo{journal}{Astrophysical Journal}
  \textbf{\bibinfo{volume}{767}}, \bibinfo{eid}{41} (\bibinfo{year}{2013}).

\bibitem[{\citenamefont{{Dresing} et~al.}(2014)\citenamefont{{Dresing},
  {G{\'o}mez-Herrero}, {Heber}, {Klassen}, {Malandraki}, {Dr{\"o}ge}, and
  {Kartavykh}}}]{Dresing_2014b}
\bibinfo{author}{\bibfnamefont{N.}~\bibnamefont{{Dresing}}},
  \bibinfo{author}{\bibfnamefont{R.}~\bibnamefont{{G{\'o}mez-Herrero}}},
  \bibinfo{author}{\bibfnamefont{B.}~\bibnamefont{{Heber}}},
  \bibinfo{author}{\bibfnamefont{A.}~\bibnamefont{{Klassen}}},
  \bibinfo{author}{\bibfnamefont{O.}~\bibnamefont{{Malandraki}}},
  \bibinfo{author}{\bibfnamefont{W.}~\bibnamefont{{Dr{\"o}ge}}},
  \bibnamefont{and}
  \bibinfo{author}{\bibfnamefont{Y.}~\bibnamefont{{Kartavykh}}},
  \bibinfo{journal}{Astronomy \& Astrophysics} \textbf{\bibinfo{volume}{567}},
  \bibinfo{eid}{A27} (\bibinfo{year}{2014}).

\end{thebibliography}
\end{document}